**Fabrication of glucose biosensors by inkjet printing**

Tianming Wang,[a] Christopher C. Cook,[a] Simona Serban[b], Tarif Ali[a], Guido Drago[b] and Brian Derby*[a]

[a] School of Materials, University of Manchester, Oxford Road., Manchester, M1 7HS, UK. *E-mail: brian.derby@manchester.ac.uk.

[b] Gwent Electronic Materials Ltd, Monmouth House, Mamhilad Park, Pontypool, Torfaen NP4 0HZ, UK.

Inkjet printing has been used to fabricate glucose sensors using glucose oxidase and screen printed carbon electrodes. By appropriate selection of printing and drying conditions we are able to fabricate sensor structures that show a good linear response to glucose concentration. In order to achieve these structures we must carefully control the spreading and drying of the enzyme solution on the carbon electrode. Carbon electrode suirfaces are hydrophobic and Triton X was used as a surfactant to allow full coverage of the electrode surface. During drying, under ambient conditions the enzyme solution segregates to form a ring deposit (coffee staining). Coffee staining is shown to be deleterious to sensor performance and it can be removed by drying in a reduced humidity environment.

**Introduction**

For many point of care (POC) applications, a cost effective and disposable biosensor is advantageous. It should typically give efficient and quick response measurements with a high degree of reliability. This implies that fabricated biosensors need to be highly uniform, have reproducible sensing elements and be manufactured at a modest cost price. Glucose biosensors are commonly used for diagnosing diabetic patients by quantifying blood glucose levels using electrodes with immobilized glucose oxidase (GOX) enzyme. Redox between the electrode surface and the active site of the enzyme generate electron transfers leading to the flow of a current that is used to measure the amount of analyte present. In order to achieve rapid and reproducible deposition of several solutions within small areas, screen printing and photolithography are currently the most widely used manufacturing methods. Screen printing holds advantage over photolithography for large scale biosensor production due to its high precision, speed and low cost. This technique forces special inks or pastes through a mesh screen to form an image onto a surface. In this way different parts of a biosensor have been produced, using electron conductive or enzyme containing pastes (Albareda-Sirvent et al. 2000), with resolution as high as 390 dpi (approximately 65 μm), and dot diameters in the range 40-70 μm. However, one drawback of screen printing is the relatively large quantity of enzyme ink that must be applied to the print screen with much of the enzyme left unused. This is not economical on the industrial scale because enzymes are relatively expensive compared with other materials used in biosensor manufacture.

Inkjet printing is a non-contact digital printing method that can be used for highly precise rapid deposition of accurately metered material (Derby 2010). It allows the deposition of small volume droplets, currently with a minimum volume of approximately 1 pl, on a large range of compatible surfaces e.g. glass, plastic or metal. Precision and reproducibility are potentially higher than screen printing, with lower formulated ink viscosities than screen printing pastes. Consequently, resolutions higher than 1200 dpi (dots per inch) are achievable with a mean dot diameter around 15 – 40 μm, depending on the extent of spreading to equilibrium on a given surface. In addition, non-contact transfer of ink from printhead to substrate reduces contamination and allows the placement of materials on contact-sensitive substrates. The advantages of inkjet printing over screen printing make mass production of smaller and cheaper biosensors a distinct possiblity.

Inkjet printing generates and accurately positions picolitre (pl) volumes of fluid. There are two forms of inkjet printing distinguished by their drop generation mechanisms, continuous inkjet printing (CIJ) and drop-on-demand inkjet printing (DOD). The differences between the two drop generation mechanisms have been reviewed in detail elsewhere (Martin et al. 2008), however it is appropriate to distinguish between them here. In CIJ, a liquid (the ink) is passed througha small diameter orifice and the resulting fluid jet breaks up into a train of identical, regularly spaced droplets through the Rayleigh instability. On generation, the droplets are charged electrically enabling them to be steered by an electric field. Drops not required for printing are deflected away from the substrate and recirculated. In DOD, drops are generated by propagating a pressure pulse in a fluid filled chamber only when a drop is needed. Drops are located in position by moving the printhead to the desired location before drop generation.

There is a further subdivision of DOD technology determined by the mechanism through which the pressure pulse is generated (Derby 2010). With thermal DOD, a microheater vaporizes a small pocket of the fluid; the formation and collapse of the vapour bubble generates the pressure pulse. In piezoelectric DOD the pulse is formed by mechanical actuation of the chamber walls. Desktop inkjet printers are usually thermal DOD but larger scale graphics printers and those used for printing functional materials are usually piezoelectric DOD.

There has been considerable work in recent years in the use of inkjet printing to produce a range of sensor structures (Dua et al. 2010; Hossain et al. 2009; Jang et al. 2007; Li et al. 2007). Most biosensor applications are likely to use piezoelectric DOD printing. CIJ is considered unsuitable because of the risk of contamination that occurs during ink recirculation (Derby 2008). Thermal printers might be expected to damage biological fluids, however only small quantities of fluid are heated during drop generation and this technique has been successfully used to deposit enzymes and cells without damage. Indeed, pioneering work by Setti has used thermal inkjet printing technology to print glucose oxidase sensor structures and demonstrate their feasibility (Setti et al. 2005). However, we believe piezoelectric printing is a conservative choice for printing biological materials because it does not significantly heat the ink during deposition (Saunders et al. 2008). With both forms of DOD printing, significant transient shear rates occur during drop generation (in excess of $10^5$ $s^{-1}$) and there have been reports that these high shear rates may damage biological materials (Di Risio and Yan 2007; Nishioka et al. 2004). In a recent study we investigated in detail whether inkjet printing affected the activity of GOX and found that there appeared to be a reduction in enzymatic activity as measured using a fluorescent assay (Cook et al. 2010). Enzyme activity was found to correlate strongly and inversely with the amplitude of the actuating pulse used to drive the piezoelectric drop generator, however a detailed study of the protein molecule using a range of light scattering methods was unable to identify any specific change in molecular size shape and secondary structure. Here we further investigate how inkjet printing influences GOX by characterising the behaviour of a simple biosensor structure produced by inkjet printing GOX solution onto carbon electrodes.

**Experimental Procedure**

GOX with specific activity at 278 U/mg and a proprietary stabilizing reagent were supplied by Applied Enzyme Technology (Product code: Q2090114D10: AET Ltd, Pontypool, UK). This was immobilized on a L-shaped silver reference electrode and carbon electrode with a $3 \times 3$ mm square working area, screen-printed on polyester sheets (product code: BE2040908D2/001: Gwent Electronic Materials Ltd.); henceforth referred to as electrode 1. Scanning electron microscopy (SEM) showed that the carbon electrode contained graphite flakes of 5-8 μm diameter. Electrodes made from the identical graphite material with 1 mm diameter circular working area, surrounded by a circular counter electrode were used to investigate the behaviour of the printed ink under different drying conditions; henceforth referred to as electrode 2.

Glucose 99%, di-potassium hydrogen phosphate, potassium chloride, potassium dihydrogen phosphate, potassium ferricyanide, all purchased from Sigma-Aldrich (Gillingham, Dorset, UK), were used to prepare the glucose test solutions. 20 and 60 mg/ml GOX solutions were prepared by adding the appropriate amount of GOX into the stabilizing reagent and stirring; 0.05 vol% Triton X100 (Sigma-Aldrich) was added as a surfactant to reduce the contact angle of the deposited drops on the electrode surface. Solution viscosity was measured using a concentric cylinder rheometer (RSIII, Brookfield, Middleboro, MA, USA). Contact angle was measured by the static sessile drop method using a dedicated instrument mainly consisiting of a platform and a CCD camera, which is connected to a PC. A commercial software application Fta32 (First Ten Angstroms, Portsmouth, VA, USA), was applied to measure the contact angle, change of drop base diameter and sessile volume.

Glucose test solutions up to 30 mM were prepared by adding glucose to deionised water containing potassium dihydrogen phosphate, potassium chloride, and di potassium hydrogen phosphate, and then adjusted to pH 7.5. Finally, 32 mg/ml. Potassium ferricyanide was added as mediator. Printing trials were carried out using an in-house designed and built laboratory scale inkjet printing platform (MPP 1000) using an *x-y* table that has a positional accuracy of 3 μm (Micromech Systems, Braintree, UK). This is equipped with piezoelectric actuated inkjet printheads of internal diameter 60 μm (MJ-ATP-01, Microfab Technologies, Plano, TX, USA) with drive electronics (JetDrive III, Microfab) interfaced to a PC and controlled in a LabVIEW™ (National Instruments, Austin, TX, USA) environment. The printer was set to print at 60V with a pulse width of 25 μs and a rise and fall time of 3μs. A frequency of 5000Hz was used. This resulted in a mean drop volume of approximately 110 pl with a mean diameter of 65 μm. The principles used to select the actuating voltage pulse for a given fluid in a piezoelectric inkjet printer have been discussed in a previous publication (Reis et al. 2005).

The surface profile of the enzyme deposit after printing onto the electrode, and drying under different conditions, was determined using Phase Contrast Microscopy (PCM) (Microxam, Phase Shift Technology, Tucson, AR, USA). A potentiostat, CompactStat (Ivium Technologies BV., Eindhoven, The Netherlands) was used to take amperometric measurements. The potentiostat requires a two-electrode cell configuration, consisting of a reference silver electrode screen-printed on polyester and a working electrode on which the enzyme is deposited (configuration as in electrode 1). Printed electrodes were dried for 4 hours either under ambient conditions or in a desiccator at a controlled temperature of 4 °C, 25 °C, or 37 °C. Hand pipetted samples, either containing or excluding surfactant, were prepared and dried under the same conditions as controls. Electrochemical responses were measured by applying 100 μl glucose solution, in the presence of ferricyanide mediator, on the two combined electrodes at an applied potential of +0.5 V. The responses were measured after 10 s incubating time, and the current value read and recorded with different glucose concentrations after 10 s and 40 s, respectively. Each data point recorder represents an average from three repeated measurements. By measuring amperometric responses in GOX solution with different concentrations a calibration curve was plotted.

**Results and Discussion**

*Printing of GOX solutions*

The requirements of fluid properties and the influence of these and the actuation signal in DOD printing on printed drop size and velocity have been reviewed and discussed elsewhere (Derby 2010; Derby and Reis 2003; Reis et al. 2005). Printing parameters were set to maintain the ejected droplet diameter at approximately 65 μm. The key solution properties that may affect printability are viscosity and surface tension, while the former being more dominant than the latter. Manufacturers' information states that the viscosity for the printable liquid should be below 20 mPa.s. The measured viscosity of solutions with concentration 20 and 60 mg/ml, were 2.3 and 2.5 mPa.s respectively, confirming that the viscosity range is within the range of solution printability; addition of 0.05 vol% Triton X100 further lowers viscosity to around 2.0 mPa.s.

The printer delivers a defined number of drops on the electrode; this is determined by the GOX concentration and the quantity of enzyme required. In principle, amperometric biosensors measure a current generated from electron transfer through catalytic reactions between the immobilised redox enzyme and substrate molecule. There is a tunneling distance between the enzyme and the electrode, over which the generated electron needs to travel in order to reach the electrode (Habermuller et al. 2000). To obtain a rapid measurable response, the redox transformation needs to take place on the surface of the electrode. This requires sufficient immobilized enzyme uniformly spread on the working area of the electrode.

Unlike screen printing, inkjet printing utilises a non-contact delivery mechanism to precisely place ink drops, which then spread through capillary forces (Derby 2010). The final shape of the printed deposit will be governed initially by the contact angle of the liquid on the substrate and the area of contact or footprint of the resulting sessile drop (Smith et al. 2006; Stringer and Derby 2009, 2010). The carbon electrodes commonly used for enzymatic sensors are relatively hydrophobic and the initial large contact angle results in poor spreading of the enzyme solution after printing (Figure 1a and b).

To achieve a more uniform deposit of enzyme over the electrode, either a surfactant can be added to lower the contact angle or the printer can be programmed to distribute individual printed drops in a defined array across the electrode instead of at a single location (figure 1c) (Cook et al. 2009). Adding a surfactant to the solution reduces its surface tension and thereby reduces the contact angle. Triton X100 was chosen for this purpose because it is a non-ionic surfactant that is widely used for preserving the functional state of proteins and accelerating the electron transfer process (Moller and Lemaire 1993; Rusling and Forster 2003). The addition of Triton X100 significantly reduced the contact angle between the solution and the carbon electrode, as shown in figure 2a. In this study, the volume concentration of Triton X100 was optimized as 0.05.

Figure 2b shows the change in sessile drop base diameter (footprint) and volume after printing 999 drops (the maximum number per trigger) of GOX onto the surface of electrode 2. The drop volume decreases linearly with time, while base diameter increases rapidly in the first 30 seconds before becoming constant. This is consistent with the results displayed in Fig. 2a, where the contact angle shows a rapid initial decrease, associated with capillary spreading, for about 20 - 30 seconds (curve without surfactant), after which the drop volume shrinks at constant base diameter indicating contact line pinning. On examining Figures 2a and 2b, it can be seen that after 20 s the solution reached a stable base diameter of approximately 0.47 mm and a contact angle of 41° before the visible drop volume decreased gradually by drying or absorption with the contact diameter pinned. The linear rate of decrease of drop volume with time is consistent with simple models for the loss of fluid from an evaporating drop with a pinned contact line (Schoenfeld et al. 2008). The influence of the surfactant is to reduce the initial spreading time and decrease the initial contact angle.

*Optimizing drying conditions*

Inkjet delivery requires the material to be in liquid form; thus drying is an important part of the manufacturing process. It is well known that significant solute segregation and redistribution can occur during drying of droplets with, in the worst case, all solute depositing as a ring at the edge of the drying drop because of outward radial flows generated within the fluid drop during drying – otherwise known as the coffee stain effect (Deegan et al. 1997). PCM was used to investigate the profile of dried GOX deposits after printing on the carbon electrodes under different drying conditions. Each electrode was printed with a stabilised 60 mg/ml GOX solution, including 0.05 vol.% Triton X100, at a frequency of 5 kHz. Two runs, each of 999 drops, were deposited onto the surface of electrode 1 in order to give approximately 2 U enzyme on the surface of the electrode. These were then dried at three different temperatures a) 4 $^{o}$C, b) 25 $^{o}$C and c) 37 $^{o}$C for 24 hours, either under ambient atmospheric conditions or in a desiccator.

On drying under ambient atmospheric conditions a characteristing ring deposit formed at all drying temperatures (figure 3a). Figure 3b - d illustrates the profile of the printed deposits after drying in the desiccator. The sample dried at 25 °C in the desiccator showed no coffee staining although extensive cracking of the dried enzyme deposit is visible in figure 3c. This is probably caused by shrinking induced stresses that often occur during the drying of films.

It is interesting to note that although coffee staining is surpressed through drying in a desiccator at 25 °C, it is evident when the enzyme solution is dried at a lower temperature (figure 3b) and a higher temperature (figure 3d). Another feature that changes slightly with the different drying conditions is the total extent of drop spreading. The deposit dried under ambient humidity is seen to spread to cover the full area of the 1 mm diameter surface of electrode 2, wheras trhose dried in the desiccator showed spreading to about 500 μm at 4 °C and 25 °C, with possibly a slightly greater extent of spreading at 37 °C. This may be the result of a more rapid reduction in drop volume with time through solvent evaporation, leading to a significantly smaller drop size at the end of the capillary spreading process when dried in the desiccator. However, earlier work on the spreading of inkjet printed drops of different fluids on porous substrates has reported the spreading time to be significantly shorter than that required for capillary infiltration or evaporation (Holman et al. 2002; Wang et al. 2008b).

Coffee staining is observed with both the air dried and desicator dried specimens; although it is substantially reduced when dried at 25 °C in the desiccator. Coffee staining has been discussed in detail in a number of previous publications. Deegan proposed that the retreating contact line in a drying drop is pinned and that internal fluid flow is required to replenish fluid to the drop edges, where a greater proportion of the solvent is lost because the proximity of a large dry surface (Deegan et al. 1997, 2000). Fischer developed the initial models of Deegan et al to include a more physically realistic model of solute loss through evaporation at the contact line (Fischer 2002) and demonstrated that radial fluid flows that cause coffee staining occur during the evaporation of drops even when fluid loss is uniform across the drop. It has been suggested that Marangoni flows associated with small temperature differences that occur during drying should oppose coffee staining and that these will be greater at low temperatures (Hu and Larson 2006). However, our results in figure 4 show an increase in coffee staining as the temperature is reduced and a minimum in coffee staining at 25°C, with an enhanced effect at both higher an lower temperatures.

The difference in behaviour observed between the two drying conditions is difficult to explain through changes in the evaporation behaviour of the liquid with temperature. One possible difference between the drying of our drops on carbon electrodes is the relative rate of drop volume decrease by evaporation and absorption into the porous carbon electrode. An earlier study of the drying of nanoparticle suspensions on porous substrates found that coffee staining was enhanced on a porous surface when compared to the behaviour of an identical fluid on a solid surface (Dou and Derby 2012). It was proposed that coffee stains form when fluid removal occurs either by evaporation or absorption, as long as the contact line is pinned. It was further demonstrated that as the temperature decreases, there is a critical temperature below which draining dominates. We do not have sufficient data to accurately determine the relative rate of capillary infiltration into the carbon electrode and volume loss due to evaporation. Although calculations presented in earlier work indicate that for aqueous solutions these will be of the same order (Dou and Derby 2012).

Thus it is proosed that at 25 °C coffee staining in the desiccator is suppressed by a countervailing Marangoni flow (Deegan et al. 1997, 2000; Fischer 2002). The surface tension gradients that induce the flow opposing coffee staining are created by local temperature differences or composition differences created when solvent is lost through evaporation. On a porous surface, draining does not occur through a phase change, hence there is no mechanism to generate local composition or temperature changes within the draining fluid and thus no mechanism to oppose coffee staining (Dou and Derby 2012). The reappearance of coffee staining that is observed when the drop is cooled to 4° C, is the effect of fluid flow being predominently driven by draining. There is also evidence in the literature for a similar coffee staining behaviour occuring when protein solutions are deposited on a porous film substrate.(Wang et al. 2008a)

*Electrochemical Performance*

Preliminary electrochemical characterisation on the samples printed on electrode 2 was used to distinguish between the desiccator drying conditions. Figure 4a shows the current as a function of glucose concemmtration for sensors fabricated with 2000 drops of a 20mg/ml GOX ink on electrode 2. The 4 °C specimen shows no signal until the glucose concentration exceeds 4 mM. The specimen dried at 25 °C shows the most linear response to glucose concentration with the least scatter amongst the data. Finally the specimen dried at 37 °C shows considerable scatter at each glucose concentration and poor linearity. These results indicate the deleterious effect of coffee staining on the performance of the printed sensor. Electrode 2 was used for subsequent sensor fabrication with an enzyme concentration of 60 mg/ml. The number of drops printed was adjusted to give enzyme dosage of 1 U or 3 U. Figure 4b shows the result of chronoamperometric meaurements of inkjet printed electrodes versus reaction time at 4 different concentrations of glucose solution.

In order to characterise the printed sensor, its performance was compared against a control specimen produced by hand pipetting the required quantity of enzyme onto the electrode. In both cases a ferricyanide mediator was used. Inkjet printing doses the electrode in integral units of about 100 pl with a GOX solution concentration of 60 mg/ml. In order to produce equivalent dosing using pipette delivery a solution concentration of 6 mg/ml was used. Current responses were read and noted after the preliminary incubation period of 10 and 40 seconds, respectively, using a constant voltage. By repeating this test for each glucose concentration it is possible to construct a calibration curve of current response against glucose concentration (Figure 5). In order to determine the minimum quantity of enzyme required on the electrode, two sensor structures were printed with 1 U and 3 U of GOX. It is clear from figure 8b that the 3 U electrodes give a good linear response with change in glucose concentration after both 10 s and 40 s exposure. In contrast the 1 U sample returns a low signal with poor linearity. Comparing figure 5a with figure 5b, we can see that the inkjet printed electrodes show a stronger and more linear response to glucose concentration than is seen with the pipetted control samples.

**Conclusions**

We have successfully demonstrated the use of inkjet printing to produce compact glucose oxidase biosensors with adequate linear response to glucose concentrations over a physiological range of glucose concentrations. However, performance of the printed biosensor appears to be optimised when appropriate printing and drying conditions are chosen to allow a uniform distribution of GOX over the working electrode surface. Enzyme distribution is shown to be critically dependent on the choice of appropriate drying conditions to eliminate segregation of the enzyme during drying – coffee staining.

The contact angle of inkjet printed drops were significantly lowered using a surfactant Triton X100 which allowed better spreading of the enzyme solution over the working area of screen printed carbon electrodes. We found ideal and optimal drying conditions for inkjet printed electrodes to be at room temperature (25 $^o$C) in a desiccator, this produced the most even distribution of enzyme over the surface of the electrode with minimal coffee staining effect. As expected higher concentrations of glucose solution gave greater current responses. The electrochemical characterization of the printed sensors showed a linear relationship with glucose concentration up to 25 mM, this is an adequate range for practical clinical glucose sensing applications.

These results demonstrate that inkjet printing technology makes feasible the manufacturing of a low cost high performance single-use disposable enzymatic biosensor.

**Acknowledgments**

This work was funded through the TSB under project MNT212. CCC would like to acknowledge the support of the EPSRC and Xaar plc through the Engineering Doctorate Scheme.

# References

Albareda-Sirvent, M., Merkoci, A., Alegret, S., 2000. Configurations used in the design of screen-printed enzymatic biosensors. A review. Sensor. Actuat. B-Chem.69(1-2), 153-163.

Cook, C., Wang, T., Derby, B., 2009. Inkjet Printing of Enzymes for Glucose Biosensors. In: Murthy, S.K., Khan, S.A., Ugaz, V.M., Zeringue, H.C. (Eds.), Materials and Strategies for Lab-on-a-Chip - Biological Analysis, Cell-Material Interfaces and Fluidic Assembly of Nanostructures, pp. 103-109.

Cook, C., Wang, T., Derby, B., 2010. Inkjet delivery of glucose oxidase. Chem. Comm. 46(30), 5452-5454.

Deegan, R.D., Bakajin, O., Dupont, T.F., Huber, G., Nagel, S.R., Witten, T.A., 1997. Capillary flow as the cause of ring stains from dried liquid drops. Nature 389(6653), 827-829.

Deegan, R.D., Bakajin, O., Dupont, T.F., Huber, G., Nagel, S.R., Witten, T.A., 2000. Contact line deposits in an evaporating drop. Phys. Rev. E 62(1), 756-765.

Derby, B., 2008. Bioprinting: inkjet printing proteins and hybrid cell-containing materials and structures. J. Mater. Chem. 18(47), 5717-5721.

Derby, B., 2010. Inkjet Printing of Functional and Structural Materials: Fluid Property Requirements, Feature Stability, and Resolution. Annual Rev. Mater. Res., Vol 40, pp. 395-414.

Derby, B., Reis, N., 2003. Inkjet printing of highly loaded particulate suspensions. MRS Bull. 28(11), 815-818.

Di Risio, S., Yan, N., 2007. Piezoelectric ink-jet printing of horseradish peroxidase: Effect of ink viscosity modifiers on activity. Macromol. Rap. Comm. 28(18-19), 1934-1940.

Dou, R., Derby, B., 2012. Formation of Coffee Stains on Porous Surfaces. Langmuir 28(12), 5331-5338.

Dua, V., Surwade, S.P., Ammu, S., Agnihotra, S.R., Jain, S., Roberts, K.E., Park, S., Ruoff, R.S., Manohar, S.K., 2010. All-Organic Vapor Sensor Using Inkjet-Printed Reduced Graphene Oxide. Angew. Chem. Inter. Edit. 49(12), 2154-2157.

Fischer, B.J., 2002. Particle convection in an evaporating colloidal droplet. Langmuir 18(1), 60-67.

Habermuller, L., Mosbach, M., Schuhmann, W., 2000. Electron-transfer mechanisms in amperometric biosensors. Fresen. J. Anal. Chem. 366(6-7), 560-568.

Holman, R.K., Cima, M.J., Uhland, S.A., Sachs, E., 2002. Spreading and infiltration of inkjet-printed polymer solution droplets on a porous substrate. J. Colloid Interf. Sci. 249(2), 432-440.

Hossain, S.M.Z., Luckham, R.E., Smith, A.M., Lebert, J.M., Davies, L.M., Pelton, R.H., Filipe, C.D.M., Brennan, J.D., 2009. Development of a Bioactive Paper Sensor for Detection of Neurotoxins Using Piezoelectric Inkjet Printing of Sol-Gel-Derived Bioinks. Anal. Chem. 81(13), 5474-5483.

Hu, H., Larson, R.G., 2006. Marangoni effect reverses coffee-ring depositions. J. Phys. Chem. B 110(14), 7090-7094.

Jang, J., Ha, J., Cho, J., 2007. Fabrication of water-dispersible polyaniline-poly(4-styrenesulfonate) nanoparticles for inkjet-printed chemical-sensor applications. Adv. Mater. 19(13), 1772-+.

Li, B., Santhanam, S., Schultz, L., Jeffries-El, M., Iovu, M.C., Sauve, G., Cooper, J., Zhang, R., Revelli, J.C., Kusne, A.G., Snyder, J.L., Kowalewski, T., Weiss, L.E., McCullough, R.D., Fedder, G.K., Lambeth, D.N., 2007. Inkjet printed chemical sensor array based on polythiophene conductive polymers. Sensor. Actuat. B-Chem. 123(2), 651-660.

Martin, G., D., Hoath, S.D., Hutchings, I.M., 2008. Inkjet printing - the physics of manipulating liquid jets and drops. J. Phys. Conf. Ser. 105, 012001.

Moller, J.V., Lemaire, M., 1993. Detergent binding as a measure of hyprophobicity surface-area of integral membrane proteins. J. Biol. Chem. 268(25), 18659-18672.

Nishioka, G.M., Markey, A.A., Holloway, C.K., 2004. Protein damage in drop-on-demand printers. JACS 126(50), 16320-16321.

Reis, N., Ainsley, C., Derby, B., 2005. Ink-jet delivery of particle suspensions by piezoelectric droplet ejectors. J. Appl. Phys. 97(9).

Rusling, J.F., Forster, R.J., 2003. Electrochemical catalysis with redox polymer and polyion-protein films. J. Colloid Interf. Sci. 262(1), 1-15.

Saunders, R.E., Gough, J.E., Derby, B., 2008. Delivery of human fibroblast cells by piezoelectric drop-on-demand inkjet printing. Biomaterials 29(2), 193-203.

Schoenfeld, F., Graf, K., Hardt, S., Butt, H.-J., 2008. Evaporation dynamics of sessile liquid drops in still air with constant contact radius. Inter. J. Heat Mass Tran. 51(13-14), 3696-3699.

Setti, L., Fraleoni-Morgera, A., Ballarin, B., Filippini, A., Frascaro, D., Piana, C., 2005. An amperometric glucose biosensor prototype fabricated by thermal inkjet printing. Biosens. Bioelectron/ 20(10), 2019-2026.

Smith, P.J., Shin, D.Y., Stringer, J.E., Derby, B., Reis, N., 2006. Direct ink-jet printing and low temperature conversion of conductive silver patterns. J. Mater. Sci. 41(13), 4153-4158.

Stringer, J., Derby, B., 2009. Limits to feature size and resolution in inkjet printing. J. Europ. Ceram. Soc. 29(5), 913-918.

Stringer, J., Derby, B., 2010. Formation and Stability of Lines Produced by Inkjet Printing. Langmuir 26(12), 10365-10372.

Wang, J., Thom, V., Hollas, M., Johannsmann, D., 2008a. Dye deposition patterns obtained in line printing on macroporous membranes: Improvement of line sharpness by liquid redistribution. J. Memb. Sci. 318(1-2), 280-287.

Wang, T.M., Patel, R., Derby, B., 2008b. Manufacture of 3-dimensional objects by reactive inkjet printing. Soft Matter 4(12), 2513-2518.

**Figure Captions**

**Figure 1** a) Sessile drop after printing 4000 drops of GOX solution onto a carbon electrode 1. b) Plan view of drop in (a). c) Plan view of 20 x 20 grid printed on electrode 1, with 10 drops printed at each grid location.

**Figure 2** Time evolution of a sessile drop of the GOX solution placed on a carbon electrode surface. a) Change of the contact angle as a function of time, with and without Triton X100. b) Change of base diameter and sessile volume of 999 drops GOX solution printed onto the surface of electrode 2.

**Figure 3** PCM images of GOX deposit after printing and drying on 1 mm diameter electrodes under the following drying conditions: a) air, 4°C; b) desiccator, 4°C; c) desiccator, 25°C and d) desiccator, 37°C.

**Figure 4** a) Electrochemical current response to glucose solutions of fabricated electrodes dried at 4 $^o$C, 25 $^o$C (RTP) and 37 $^o$C in a desiccator. b) Chronoampemetric response of inkjet printed electrodes as a function of reaction time with various glucose concentrations; GOX 3 U was printed from a 60 mg/ml suspension.

**Figure 5** Sensor response as a function of glucose concentration for samples prepared by (a) hand pipetting, and (b) inkjet printing.

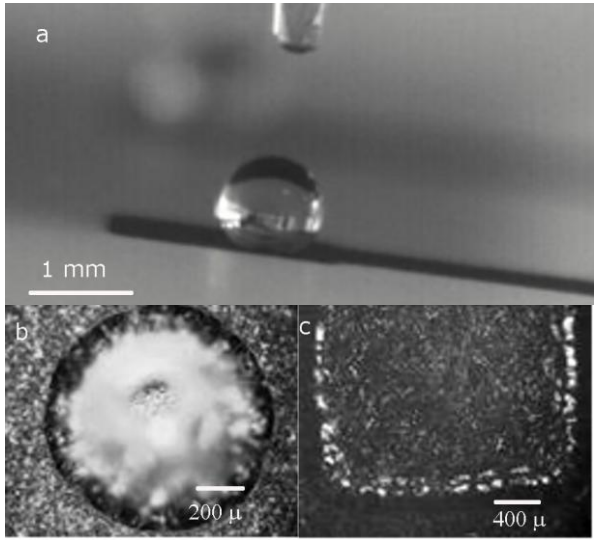

**Figure 1** a) Sessile drop after printing 4000 drops of GOX solution onto a carbon electrode. b) Plan view of drop in (a). c) Plan view of 20 x 20 grid with 10 drops printed at each grid location.

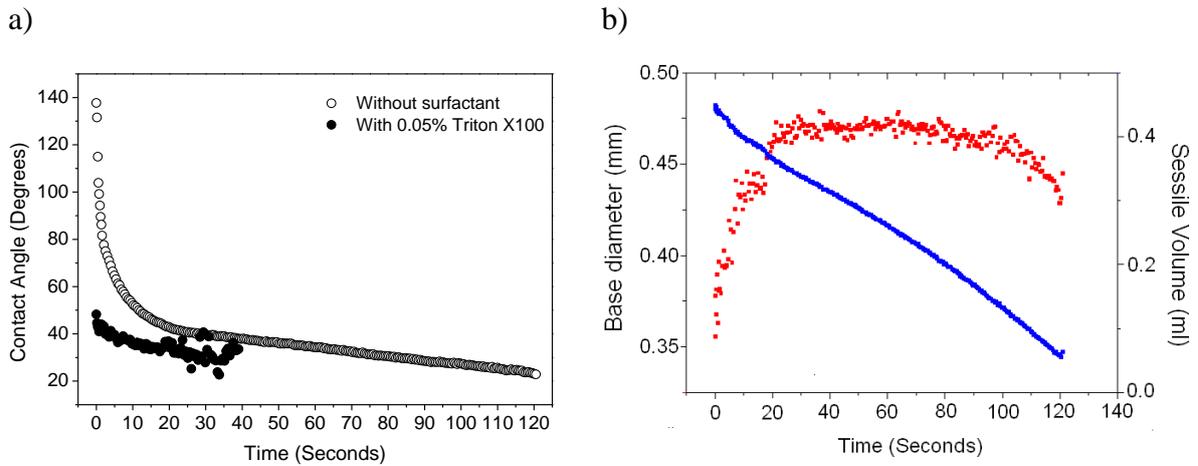

**Figure 2** Time evolution of a sessile drop of the GOX solution placed on a carbon electrode surface. a) Change of the contact angle as a function of time, with and without Triton X100. b) Change of base diameter and sessile volume of 999 drops GOX solution printed onto the surface of electrode 2.

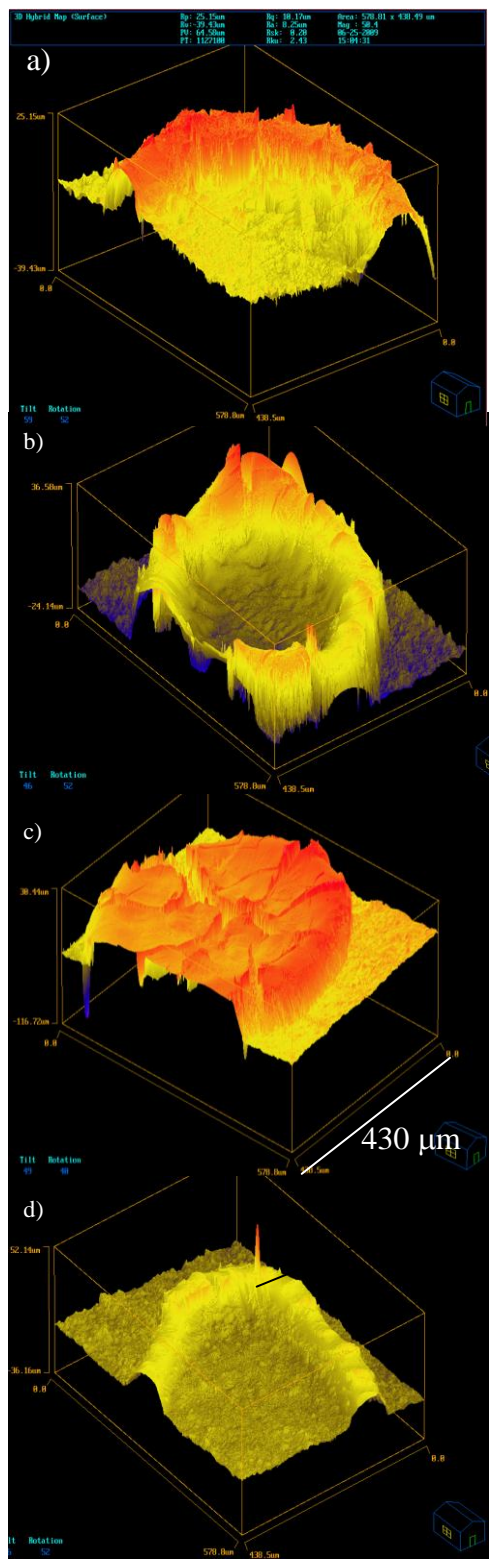

**Figure 3** PCM images of the enzyme printed on 1 mm diameter electrodes after drying: a) dried in air at 4°C; b) dried in a dessicator at 4°C; c) dried in a dessicator at 25°C and d) dried in a dessicator at 37°C.

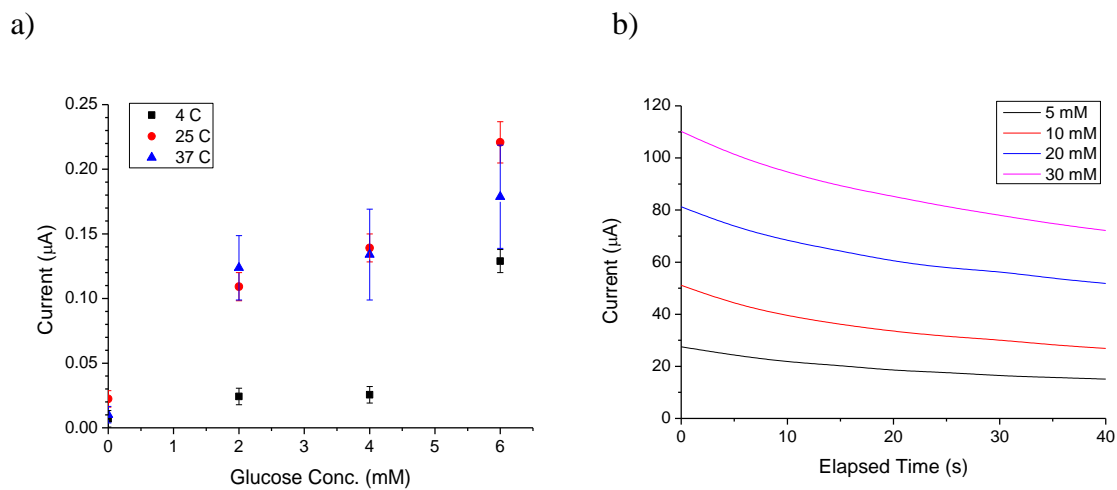

**Figure 4** a) Electrochemical current response to glucose solutions of fabricated electrodes dried at 4 °C, 25 °C (RTP) and 37 °C in a desiccator. b) Chronoampemetric response of inkjet printed electrodes as a function of reaction time with various glucose concentrations; GOX 3 U was printed from a 60 mg/ml suspension.

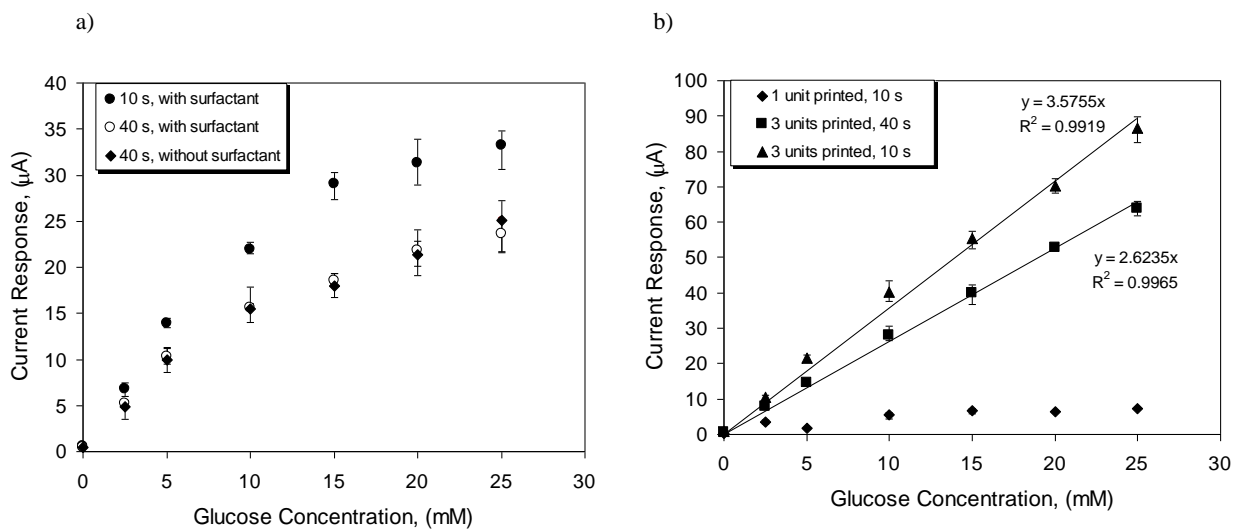

**Figure 5** Sensor response as a function of glucose concentration for samples prepared by (a) hand pipetting, and (b) inkjet printing.